\documentclass[12pt,aps,ams,amsfonts,nofootinbib]{revtex4}

\usepackage[usenames,dvipsnames]{color}
\usepackage[utf8]{inputenc}
\usepackage{lipsum}
\usepackage[sc]{mathpazo}
\usepackage{graphicx}
\usepackage{amssymb}
\usepackage{subfigure}
\usepackage{stackrel}
\usepackage{enumitem}
\usepackage{amsmath}
\usepackage{amsfonts}
\usepackage{bm}
\usepackage{color}
\usepackage{verbatim} 
\usepackage{tcolorbox}
\usepackage[normalem]{ulem}
\usepackage{hyperref}
\usepackage{soul}

\usepackage[usenames,dvipsnames]{color}
\usepackage{amsfonts}

\newlist{subquestion}{enumerate}{1}
\setlist[subquestion,1]{label=(\alph*)}

\newcommand{\ket}[1]{\ensuremath{\left|{#1}\right\rangle}}

\newcommand{\ojo}[1]{#1}

\newcommand{\beq}{\begin{equation}}
	\newcommand{\eeq}{\end{equation}}
\newcommand{\bse}{\begin{subequations}}
	\newcommand{\ese}{\end{subequations}}\newcommand{\bea}{\begin{eqnarray}}
	\newcommand{\eea}{\end{eqnarray}}
\newcommand{\bit}{\begin{itemize}}
	\newcommand{\eit}{\end{itemize}}
\newcommand{\bpmatrix}{\begin{pmatrix}}
	\newcommand{\epmatrix}{\end{pmatrix}}

\newcommand{\be}{\begin{equation}}
	\newcommand{\ee}{\end{equation}}
\newcommand{\ben}{\begin{eqnarray}}
	\newcommand{\een}{\end{eqnarray}}

\begin{document}

\title[ESD in fermionic systems under collective decoherence]{Sudden death of entanglement in fermionic systems under collective decoherence}

\author{D. G. Bussandri$^{1}$, A. P. Majtey$^{2,3}$ and A. Valdés-Hernández$^4$}
\affiliation{$^1$Instituto  de  F\'isica  La  Plata  (IFLP),  CONICET,  and  Departamento  de  F\'isica,  Facultad  de Ciencias Exactas, Universidad Nacional de La Plata, C.C. 67, 1900 La Plata, Argentina}
\affiliation{$^2$Facultad de Matem\'atica, Astronom\'{\i}a, F\'{\i}sica y Computaci\'on, Universidad Nacional de C\'ordoba, Av. Medina Allende s/n, Ciudad Universitaria, X5000HUA C\'ordoba, Argentina}
\affiliation{$^3$ Instituto de F\'isica Enrique Gaviola, Consejo Nacional de Investigaciones Cient\'{i}ficas y T\'ecnicas de la Rep\'ublica Argentina, Av. Medina Allende s/n, X5000HUA, C\'{o}rdoba, Argentina}
\affiliation{$^4$ Instituto de F\'{\i}sica, Universidad Nacional Aut\'{o}noma de M\'{e}xico, Apartado Postal 20-364, Ciudad de M\'{e}xico, Mexico}

\begin{abstract}
We analyze the dynamics of entanglement due to decoherence in a system of two identical fermions with spin $3/2$ interacting with a global bosonic environment. We resort to an appropriate  measure of the so-called fermionic entanglement to quantify the fermionic correlations, and compare its dynamics with that of a pair of distinguishable qubits immersed in the same environment. According to the system's initial state, three types of qualitatively different dynamics are identified: i) \textit{invariant regime}, corresponding to initial states that belong to a decoherence free subspace (DFS), which maintain invariant their entanglement and coherence throughout the evolution; ii) \textit{exponential decay}, corresponding to initial states orthogonal to the DFS, and evolve towards states whose entanglement and coherence decrease exponentially; iii) \textit{entanglement sudden death}, corresponding to initial states that have some overlap with the DFS and exhibit a richer dynamics leading, in particular, to the sudden death of the fermionic entanglement, while the coherence decays exponentially.  
Our analysis offers insights into the dynamics of entanglement in open systems of identical particles, into its comparison with the distinguishable-party case, and into the existence of decoherence free subspaces and entanglement sudden death in indistinguishable-fermion systems. 
\end{abstract} 
\maketitle

\section{Introduction\label{sec:intro}}

Entanglement and coherence are fundamental features of quantum information
processing (QIP) \cite{Horodecki2009,Streltsov2017}. The former, exhibiting non-classical correlations, is commonly referred to as a key resource
for quantum information tasks \cite{Theurer2017}. The latter, corresponding to the capability of a
system to allow for the interference of its possible states, has two important roles both in quantum control schemes and in settling the conditions to allow for
quantum correlations, included entanglement \cite{Adesso2016}. 

In real experimental
applications, the coupling of a system to the surrounding environment generally causes decoherence, which manifests as a loss of coherence and entanglement and leads to the concomitant loss of the quantum properties of the system (an algorithm based on systems whose states are completely decohered
can be simulated by classical computers \cite{Ekert1998}). Consequently, the quantum information
community is continuously researching around a specific goal: to
avoid decoherence in a particular quantum information task. A
possible solution is implemented by error correction theory \cite{Divincenzo2016}. An alternative
scheme consists in avoiding errors, by encoding the information
employing states belonging to decoherence-free subspaces (DFS) \cite{Duan1998,Duan1998v2}. A remarkable
model within this latter approach is \emph{collective decoherence}, which implies
that the environment couples in the same way to each part of the quantum
global system \cite{DeFilippo2000,Yang2001}. There exist experimental applications implementing
this approach to different systems such as trapped ions, superconducting
devices and neutral atoms \cite{Aolita2007,Brown2003,Cen2006,Zhang2006,Wu2008,Brion2007}.

As for entanglement, a typical behaviour when the (entangled) system
couples to its environment is an exponential decay. However, under certain
quantum channels it has been shown that the entanglement may completely
vanish in a finite time, dynamic denominated as \emph{entanglement sudden death}
(ESD) \cite{Wang2018,Sharma2004}. This notable phenomenon has been the subject of several theoretical and experimental studies and is recognized as more disruptive (when compared with the typical decay) to QIP due to the complete disappearance of entanglement \cite{Yu2004,Yu2006,Yu2009,AlQasimi,Weinstein,BLCB09,Man,Mazzola2009}. Thus, it is important to identify the possible initial set-ups and interactions which can exhibit ESD .

Many physical applications in QIP involve systems
composed of identical particles \cite{ESBL02}, however, in contrast to what happens with distinguishable-particle systems, less attention has been paid to the dynamics of entanglement and coherence in these composite systems with exchange symmetry. A first important difference is that the model of local interaction with independent reservoirs
ceases to be valid when dealing with identical, indistinguishable parties. Indeed, an open bipartite system preserves the symmetry under the exchange of its (non-interacting) parts if and only if there is a common environment, so that the evolution is global (non-local) \cite{VHMP2015}. This is especially relevant when studying
decoherence processes in identical-particle systems, a matter that is analyzed below, in relation with systems of two identical fermions \cite{Nizama2019}.

In this paper, we study the collective decoherence approach in a system of two indistinguishable fermions, and compare the evolution of both the (fermionic) entanglement and the coherence, with that of a two-distinguishable-qubit system embedded in a global environment. We find conditions that guarantee the existence of a DFS when considering two
identical fermions under collective and non-dissipative decoherence, and also the conditions leading to ESD in that same system. 

The article is organized as follows. In Sec. \ref{sec:Preliminaries} we introduce the reader to the basics of entanglement in indistinguishable-fermion systems, in particular to the corresponding (fermionic) entanglement measure, and briefly review its distinguishable-qubit counterpart. Section \ref{sec:DynamicalModel} is devoted to present the surrounding environment and the non-dissipative dynamical model under which the central system (either constituted by fermions or qubits) will evolve. In Sec. \ref{sec:Dynamics of entanglement} we consider different initial states and investigate the dynamics of the entanglement and the coherence in the open system consisting of two identical fermions and compare it with that of a two-(distinguishable)-qubit system. In particular, our examples disclose the 
existence of decoherence-free subspaces for each system of interest, and also the presence of entanglement sudden death and sudden birth, for a particular choice of initial states. We summarize and conclude this work in Sec. \ref{sec:conclusions}.

\section{Entanglement in two-identical-fermion systems} \label{sec:Preliminaries}

Let us consider a pair of identical, indistinguishable particles, and denote with $\mathcal{H}$ the single-particle Hilbert space (with $\dim \mathcal{H}=d$), with an orthonormal basis $\{\ket{i}\}=\{\ket{1}, \ket{2},\dots, \ket{d}\}$. Let $\mathcal{H}_S=\mathcal{H}\otimes\mathcal{H}$ stand for the Hilbert space of the composite (two-particle) system $S$. The antisymmetric subspace of $\mathcal{H}_S$, namely $\mathcal{H}_-$, has dimension $d(d-1)/2$, and is spanned by vectors $\{\ket{\psi^{-}_{n}}\}$ with $n\in\{1,2,\dots,d(d-1)/2\}$. In its turn, the symmetric subspace of $\mathcal{H}_S$, namely $\mathcal{H}_+$, has dimension $d(d+1)/2$, and a basis composed of the vectors $\{\ket{\psi^{+}_{k}}\}$ with $k\in\{1,2,\dots,d(d+1)/2\}$.

Writing $d=2s+1$, it is convenient to resort to an angular momentum representation \cite{ESBL02}, and to identify each state $\ket{i}$ of the single-particle basis with the angular momentum states $|s, m_s \rangle $, with $m_s\in\{-s,\dots, s\}$, so that 
\beq \label{basis1}
\{\ket{1}=\ket{s,s}, \ket{2}=\ket{s,s-1}, ..., \ket{n}=\ket{s,-s}\}. 
\eeq
Within this representation, the eigenstates $\{ |j, m\rangle \}$ (with $-j\le m
\le j$ and $0\le j\le 2s$) of the total angular momentum
operators ${J}_z$ and ${J}^2$, constitute a natural basis of $\mathcal{H}_S$. The antisymmetric eigenstates, characterized
by an even value (including $0$) of the quantum number $j$ \cite{F62,D02}, constitute a suitable basis $\{\ket{\psi^{-}_{n}}\}$ of $\mathcal{H}_-$, whereas the remaining (symmetric) states (with $j$ odd) constitute a basis $\{\ket{\psi^{+}_{k}}\}$ of $\mathcal{H}_+$. 

If $S$ represents a system composed of a pair of identical fermions, the appropriate Hilbert space for describing the system is $\mathcal{H}_-$. The antisymmetric combination (with $\ket{ij}=\ket{i}\otimes\ket{j}$)
\beq\label{slater}
|\psi^{sl}_{ij}\rangle=\frac{1}{\sqrt{2}}(\ket{ij}-\ket{ji}),\quad (i\neq j)
\eeq
is called a Slater determinant (state with Slater rank 1). The composite system of two identical fermions
is said to be separable \ojo{(or non-entangled; throughout this paper we mean entanglement between particles, as opposed to entanglement between modes)} if and only
if its density matrix can be decomposed as a statistical mixture of pure states of Slater rank 1: \cite{GMW02}
\begin{equation}
\rho^{sep}_{ff}=\sum_{ij}p_{ij}|\psi^{sl}_{ij}\rangle\langle\psi^{sl}_{ij}|
\label{rhosep},
\end{equation} 
where $p_{ij}\geq 0$, and $\sum_{ij}p_{ij}=1$. From here it follows that in order to describe entangled states of indistinguishable fermions, we need to resort to basis of $S$ that includes elements different from Slater determinants. 

Now, according to the discussion following Eq. (\ref{basis1}), for $s=1/2$ the basis $\{\ket{\psi^{-}_{n}}\}$ possess a single element (with Slater rank 1), and hence no entanglement is present. Therefore the fermion system of lowest dimensionality exhibiting the phenomenon of entanglement corresponds to $s\geqslant3/2$, or rather $d\geqslant4$ and dim$\mathcal{H}_{-}\geqslant6$. Thus, for example, for $s=3/2$ ($d=4$), only the states $|j, m \rangle$ with $j\in\{0,2\}$ are antisymmetric, and the basis of $\mathcal{H}_-$ becomes 
\beq \label{estadosf}
\{\ket{\psi^{-}_{n}}\}=\{|2, 2\rangle, |2, 1\rangle,|2, 0\rangle,|2, -1\rangle |2, -2\rangle,|0,0\rangle\}.
\eeq

Determining whether a generic two-fermion state is entangled or not is still an open problem, yet important progress has been made for some states. In particular, necessary and sufficient separability criteria have been formulated for two-fermion pure states in terms of appropriate entropic measures (see \cite{PMD09} and references therein). Moreover, for fermionic systems with $d=4$ a closed
analytical expression for the amount of entanglement, or \textit{fermionic concurrence} $C_f(\rho_{ff})$, in a general (pure or mixed) two-fermion state $\rho_{ff}$ is
known \cite{ESBL02},
\begin{equation}\label{concurrencef}
C_f(\rho_{ff})=\textrm{max}\{0,\lambda_1-\lambda_2-\lambda_3-\lambda_4-\lambda_5-\lambda_6\},
\end{equation}
\noindent
where the $\lambda_i$'s are, in decreasing order, the square roots of the eigenvalues
of $\rho_{ff}\tilde{\rho}_{ff}$ with $\tilde{\rho}_{ff}={\mathbb D}\rho_{ff} {\mathbb D}^{-1}$, and 
\begin{equation}\label{matrixD}
{\mathbb D}=\left(
\begin{array}{cccccccc}
0 & 0 & 0 & 0 & 1 & 0 \\
0 & 0 & 0 & -1 & 0 & 0 \\
0 & 0 & 1 & 0 & 0 & 0 \\
0 & -1 & 0 & 0 & 0 & 0 \\
1 & 0 & 0 & 0 & 0 & 0 \\
0 & 0 & 0 & 0 & 0 & 1 \\
\end{array}
\right)\kappa,
\end{equation}

\noindent
where $\kappa$ is the complex conjugation operator. The matrix ${\mathbb D}$ is expressed
in the basis with (ordered) elements:
$|2,2\rangle$, $|2,1\rangle$, $|2,0\rangle$, $|2,-1\rangle$, $|2,-2\rangle$,
and $i|0,0\rangle$. Notice that this is not strictly the total angular momentum basis, due to the additional phase of the last element.    

Table \ref{Table} shows the concurrence (\ref{concurrencef}) for each of the states (\ref{estadosf}). The states $|0,0\rangle$ and $|2,0\rangle$ are maximally entangled, 
while all the other states in the list correspond to single Slater determinants thus 
have zero (fermionic) entanglement.

The fermionic concurrence is an extension, to identical-fermion systems,  of the usual concurrence $C_q$ which for a two-(distinguishable)-qubit mixed state $\rho_{qq}$ is given by \cite{Wootters}
\begin{equation}\label{concurrence}
C_q(\rho_{qq})=\textrm{max}\{0,\lambda_1-\lambda_2-\lambda_3-\lambda_4\},
\end{equation}
\noindent
where the $\lambda_i$'s are, in decreasing order, the square roots of the eigenvalues
of $\rho_{qq}\tilde{\rho}_{qq}$ with $\tilde{\rho}_{qq}=(\sigma_y\otimes\sigma_y)\rho^*_{qq} (\sigma_y\otimes\sigma_y)$, and the complex conjugation is taken in the computational basis
\be \label{comp}
\{\ket{k}\}=\{\ket{00}, \ket{01},\ket{10},\ket{11}\},
\ee
with $\sigma_z\ket{0}=\ket{0}$, and $\sigma_z\ket{1}=-\ket{1}$. From here it follows that the elements of $\{\ket{k}\}$ are eigenstates of the total angular momentum along the $\boldsymbol{\hat z}$ direction $J_z=\tfrac{1}{2}\sigma_z\otimes\mathbb I_2+\mathbb I_2\otimes\tfrac{1}{2}\sigma_z$, with $\mathbb I_2$ the $2\times2$ identity operator (throughout the paper we put $\hbar=1$).

\begin{center}
	\begin{table}
		\begin{center}
			\ojo{
				\begin{tabular}{l|c}
					&$C_f$\\ \hline
					$\ket{\psi^{-}_{1}}=|2,2\rangle = |\psi^{sl}_{12}\rangle $ & 0\\
					$\ket{\psi^{-}_{2}}=|2,1\rangle =  |\psi^{sl}_{13}\rangle $ & 0\\
					$\ket{\psi^{-}_{3}}=|2,0\rangle = \frac{1}{\sqrt{2}}(|\psi^{sl}_{14}\rangle-|\psi^{sl}_{23}\rangle)$ & 1 \\
					$\ket{\psi^{-}_{4}}=|2,- 1\rangle =  |\psi^{sl}_{24}\rangle $ & 0\\
					$\ket{\psi^{-}_{5}}=|2,- 2\rangle =  |\psi^{sl}_{34}\rangle $ & 0\\
					$\ket{\psi^{-}_{6}}=|0,0\rangle = \frac{1}{\sqrt{2}}(|\psi^{sl}_{14}\rangle+|\psi^{sl}_{23}\rangle)$ & 1 
			\end{tabular}}
			\caption{Vector basis (\ref{estadosf}) with their corresponding fermionic concurrences.}\label{Table}
		\end{center}
	\end{table}
\end{center}

\section{Dynamical model}\label{sec:DynamicalModel}
\noindent
We now present the dynamical model that will be considered for analyzing the entanglement dynamics of an open
system $S$ (which can be composed of several parties). We assume that the system interacts globally with an environment $E$, under a \emph{nondissipative} interaction. That is, if the total Hamiltonian writes as
\beq
H=H_{S}+H_E+H_I,
\eeq
where $H_S$ and $H_E$ stand for the free Hamiltonians of $S$ and $E$, respectively, and $H_I$ denotes the interaction Hamiltonian, then we will focus on those interactions for which
\beq\label{conserv}
[H_{S},H]=0.
\eeq
This means that no energy exchange occurs between $S$ and $E$, so that $H_S$ is conserved. 

Further, we will be interested in those cases in which $S$ and $E$ are initially uncorrelated, so that the initial state of the complete system $S+E$ is 
\beq\label{initial}
\rho(0)=\rho_{S}(0)\otimes\rho_{E}(0).
\eeq 
The state at any time $t$ is thus 
\beq\label{evol}
\rho(t)=e^{-iHt}\big[\rho_{S}(0)\otimes\rho_{E}(0)\big]e^{iHt},
\eeq
and hence the (reduced) subsystem $S$ evolves as
\begin{eqnarray}\label{evolF}
\rho_S(t)&=&\textrm{Tr}_{E}\,\big(e^{-iHt}\big[\rho_{S}(0)\otimes\rho_{E}(0)\big]e^{iHt}\big),
\end{eqnarray}
where $\textrm{Tr}_{E}$ denotes the partial trace over the degrees of freedom of $E$.

In particular, following Privman \cite{Priv}, we will assume that the environment is represented by a bosonic bath, whose modes $\{k\}$ are characterized by creation ($a_k^{\dag}$) and annihilation ($a_k$) operators satisfying $[a_k,a_k^{\dag}]=1$.  Further, \ojo{we will consider a paradigmatic Hamiltonian of the form 
	\begin{equation} \label{Hamiltonian}
	H=H_S+H_I+H_R,
	\end{equation}
	being $H_R=\sum_k\omega_ka_k^{\dag}a_k$, the internal Hamiltonian of the bath, and $H_I=\Lambda_S\sum_k(g^*_ka_k+g_ka_k^{\dag})$, the interaction term between the bath and the system. Besides, $\omega_k$ stands for the frequency of the corresponding bath oscillator, $g_k$ is a coupling constant, and $\Lambda_S$ (which represents the pointer observable of the system $S$)} satisfies
\beq
[H_{S},\Lambda_S]=0,
\eeq
by virtue of Eq. (\ref{conserv}). This latter expression determines the basis of $\mathcal{H}_S$ that will be used in what follows, given by the common eigenstates of $H_{S}$ and $\Lambda_S$, denoted as $\{\ket{n}\}$ and satisfying
\beq \label{ketn}
H_{S}\ket{n}=E_n\ket{n},\quad\Lambda_S\ket{n}=L_n\ket{n}.
\eeq
Notice that for $H_{S}=\omega_0\sigma_z/2$ and $\Lambda_S=\sigma_z/2$, the Hamiltonian (\ref{Hamiltonian}) correspond to the well-known spin-boson model.

With the aid of the above equations, we get for the matrix elements of $\rho_S(t)$ in this basis \cite{Priv}:  
\begin{align}\label{evolFnm}
\rho^{S}_{mn}(t)&\equiv\langle m|\rho_S(t)|n\rangle\nonumber\\
&=\rho^{S}_{mn}(0)e^{i(E_n-E_m)t}\,\textrm{Tr}\,\big[e^{-iH_mt}\rho_{E}(0)e^{iH_nt}\big],
\end{align}
where $H_l$ is defined as the operator 
\beq\label{h}
H_l= \sum_kh_{lk},\quad h_{lk}=\omega_ka_k^{\dag}a_k
+L_l(g^*_ka_k+g_ka_k^{\dag}).
\eeq

In order to go further with Eq. (\ref{evolFnm}), we assume that all the modes of the bath are initially uncorrelated, so that $\rho_E(0)$ factorizes into
\beq\label{RoE}
\rho_E(0)=\Pi_k\rho_k,
\eeq
with $\rho_k$ the density matrix of the $k$-th mode. If, for example, $\rho_E(0)$ were a thermal state we would have
\beq\label{thermal}
\rho_k=Z^{-1}_ke^{-\beta\omega_ka_k^{\dag}a_k},\quad Z_k=(1-e^{-\beta\omega_k})^{-1},
\eeq
where $\beta=(k_BT)^{-1}=1/T$ (in what follows we put Boltzmann constant $k_B$ equal to $1$). Moreover, since the creation and annihilation operators for different modes commute, the trace term in Eq. (\ref{evolFnm}) rewrites as
\begin{eqnarray}\label{Tr1}
\textrm{Tr}\,\big[e^{-iH_mt}\rho_{E}(0)e^{iH_nt}\big]&=&\textrm{Tr}\,\big(\Pi_ke^{-ih_{mk}t}\rho_ke^{ih_{nk}t}\big)\\
&=&\sum_{\alpha}\langle\alpha|\big(\Pi_ke^{-ih_{mk}t}\rho_ke^{ih_{nk}t}\big)|\alpha\rangle \nonumber,
\end{eqnarray}
with $\{\ket{\alpha}\}$ an arbitrary orthonormal basis of $\mathcal{H}_E$. Taking $\ket{\alpha}=\Pi_k\ket{\alpha_k}$ with $\{\ket{\alpha_k}\}$ a basis of the $k$-th mode subsystem, we get 
\beq\label{Tr2}
\textrm{Tr}\,\big[e^{-iH_mt}\rho_{E}(0)e^{iH_nt}\big]=\Pi_k\big[\textrm{Tr}_k\,(e^{-ih_{mk}t}\rho_ke^{ih_{nk}t})\big].
\eeq

Resorting to the coherent-state representation, the trace over the (single) $k$-th mode in the right-hand-side of this equation has been calculated (for $\rho_k$ given by Eq. (\ref{RoE})) in \cite{Priv}, obtaining
\beq\label{Tr3}
\textrm{Tr}_k\,(e^{-ih_{mk}t}\rho_ke^{ih_{nk}t})=\exp(-\omega^{-2}_k|g_k|^2P_{mn,k}),
\eeq
with
\begin{eqnarray}\label{Pmn}
P_{mn,k}&=&2(L_m-L_n)^2\sin^2\frac{\omega_kt}{2}\coth\frac{\beta\omega_k}{2}+\nonumber\\
&&+i(L^2_m-L^2_n)(\sin\omega_kt-\omega_kt).
\end{eqnarray}
Gathering results, Eq. (\ref{evolFnm}) becomes
\begin{eqnarray}\label{evolFnm2}
\rho^{S}_{mn}(t)&=&\rho^{S}_{mn}(0)e^{i(E_n-E_m)t}f_{mn}(t)\nonumber\\
&=&\rho^{S}_{mn}(t)|_{g_k=0}\,f_{mn}(t),
\end{eqnarray}
where the function 
\beq\label{fmn}
f_{mn}(t)=\exp(-\sum_k\omega^{-2}_k|g_k|^2P_{mn,k})
\eeq
bears the information regarding the decoherence effects. Substituting Eq. (\ref{Pmn}) into (\ref{fmn}) we get
\beq\label{fmn2}
f_{mn}(t)=e^{-(L_m-L_n)^2\Gamma(t)}e^{-i(L^2_m-L^2_n)r(t)},
\eeq
where $r(t)=\Delta(t)-\Theta(t)$, and 
\begin{subequations}\label{defs}
	\begin{eqnarray}
	\Gamma(t)&=&2\sum_k\omega^{-2}_k|g_k|^2\sin^2\frac{\omega_kt}{2}\coth\frac{\beta\omega_k}{2},\\
	\Delta(t)&=&\sum_k\omega^{-2}_k|g_k|^2\sin\omega_kt,\\
	\Theta(t)&=&\sum_k\omega^{-2}_k|g_k|^2\omega_kt.
	\end{eqnarray}
\end{subequations}

Equations (\ref{evolFnm2}) and (\ref{fmn2}) allow us to disclose some general properties of the evolution, regardless of the specific nature of $S$. The most immediate one is that the diagonal elements $\rho^{S}_{nn}$ are not affected by the interaction, as expected for a model of decoherence without dissipation. Off-diagonal matrix elements corresponding to degenerate states with respect to $\Lambda_S$ (that is, such that $L_{m}=L_{n}$ with $n\neq m$) are also immune to the presence of the bath. 

Off-diagonal elements for which $L_{m}=-L_{n}$ are only affected by the exponential decay $e^{-(L_m-L_n)^2\Gamma(t)}=e^{-4L_n^2\Gamma(t)}$,
and those matrix terms for which $L_{m}\neq \pm L_{n}$ exhibit oscillations, in addition to exponential decay, determined by $
e^{-i(L^2_m-L^2_n)r(t)}$.
In particular, in the spin-boson model ($\Lambda_S=\sigma_z/2$) the two eigenvalues $\{L_n\}$ are $\pm(1/2)$, whence the decoherence factor in such system corresponds to exponential decay only. In fact, for any $d$-level system with $d>2$ the condition $L_{m}=-L_{n}$ cannot hold for all $n\neq m$, and consequently \emph{all} $d$-level systems ($d>2$) evolve (according to the present model) in such a way that at least one of the off-diagonal matrix elements $\rho_{nm}(t)$ is affected by the oscillating factor. 

Now, coherence measures, such as
\beq\label{coherence}
\mathcal{C}=\sum_{nm (n\neq m)}|\rho^{S}_{nm}|,
\eeq
typically involve the modulus $|\rho_{nm}|$ \cite{BCP14}, so the oscillating term of $f_{mn}$ does not play any role, and decoherence is thus manifested only through the exponencial decaying term. The oscillations that distinguish the dynamics between qubits and higher dimensional systems manifest via the (relative) phases of all $\rho^S_{nm}$.  
%

Now, coming back to Eqs. (\ref{defs}) we will assume, as is customarily done, that the bath is sufficiently large so that the density of its modes can be taken as continuous. We can thus pass from the discrete sums to the continuum, with the prescription
\beq
\sum_k|g_k|^2\rightarrow \int^{\infty}_{0}d\omega J(\omega),
\eeq
with $J(\omega)$ the spectral density. Its particular form will be assumed to be
\beq
J(\omega)=4\,J_0\,\omega\, e^{-(\omega/\omega_c)},
\eeq
with $J_0$ a dimensionless constant and $\omega_c$ the cutoff frequency, which defines the characteristic temperature $T_c=\omega_c$. Note that this choice of spectral density gives rise to a Markovian evolution, in which there are no reservoir memory effects present during the evolution \cite{Mazzola2009}. With these assumptions Eqs. (\ref{defs}) become 
\begin{subequations}\label{defsc}
	\begin{eqnarray}
	\Gamma(t)&=&\frac{J_0}{2}\int^{\infty}_0 e^{-\frac{\omega}{\omega_c}}\frac{\sin^2 (\frac{\omega t}{2})}{\omega}\coth{\Big(\frac{\omega\beta}{2}\Big)}d\omega,\\
	\Delta(t)&=&\int^{\infty}_0 e^{-\frac{\omega}{\omega_c}}\frac{\sin(\omega t)}{\omega} d\omega,\\
	\Theta(t)&=&t\int^{\infty}_0 e^{-\frac{\omega}{\omega_c}}d\omega,
	\end{eqnarray}
\end{subequations}
which simplify in the low-temperature regime, i.e., whenever $\omega_c\gg \beta^{-1}$ ($T_c\gg T$), as follows
\begin{subequations}\label{defsc2}
	\begin{eqnarray}
	\Gamma(t)&=&\frac{J_0}{8}\ln(1+\omega^2_ct^2)+\frac{J_0}{4}\ln \Big[\frac{\sinh(\pi \beta^{-1}t)}{\pi \beta^{-1}t}\Big],\\
	\Delta(t)&=&\arctan(\omega_ct),\\
	\Theta(t)&=&\omega_ct.
	\end{eqnarray}
\end{subequations}

\section{Dynamics of entanglement. Fermions vs qubits}\label{sec:Dynamics of entanglement}

\subsection{Decoherence-free subspaces}

We consider first an open system consisting of two identical fermions with a single-particle
Hilbert space of dimension 4 (equivalent to 3/2-spin fermions), initially in the state $
\rho_S(0)=\rho_{ff}(0)=|\psi_{ff}(0)\rangle\langle\psi_{ff}(0)|$,
with 
$|\psi_{ff}(0)\rangle$ a coherent superposition of the states $\{\ket{\psi^{-}_{n}}\}$. Since these are eigenstates of the total momentum operator $J_z$, by taking $H_S=\omega_0J_z$ and $\Lambda_S=J_z$ we can identify the basis $\{\ket{\psi^{-}_{n}}\}$ with the basis $\{\ket{n}\}$ satisfying Eq. (\ref{ketn}), and the eigenvalue $L_n$ with the corresponding projection eigenvalue $m$ in the angular-momentum representation (see Table \ref{Table}). Therefore we have 
\beq
L_1=-L_5=2;\quad L_2=-L_4=1;\quad L_3=L_6=0.
\eeq

According to the statements below Eqs. (\ref{defs}), this implies that the matrix elements $\rho_{nm}$ with $n,m\in\{3,6\}$ will be constant during the evolution. Consequently, by varying the coefficients in the superposition 
\beq \label{inv}
|\psi_{ff}(0)\rangle_{\textrm{inv}}=\alpha \ket{\psi^{-}_{3}}+\beta \ket{\psi^{-}_{6}},
\eeq
with $|\alpha|^2+|\beta|^2=1$, a subspace of states that are unaffected by the bath is generated. In particular, the states in the subspace maintain invariant their entanglement and their coherence, and thus constitute a DFS. For larger systems, with higher-dimensional single-particle Hilbert spaces, the decoherence-free subspace will in general become larger. 

Notice that the DFS is precisely the subspace with $m=0$. This can be understood resorting to Eq. (\ref{h}), which shows that for $L_l=0$ (in this case $m=0$) the operator $H_l$ that determines the evolution of $\rho_S$ via Eq. (\ref{evolFnm}), is the same as that in absence of interaction (observe that this is a general result, which requires only the Hamiltonian form (\ref{Hamiltonian}), and is independent of the specificities of the initial state of the bath). 

\ojo{Moreover, any state that is equivalent to (\ref{inv}) under local and exchange-symmetry-preserving transformations in $\mathcal{H}_f\otimes \mathcal{H}_f$, with $\mathcal{H}_f$ the single-fermion Hilbert space, will exhibit the same amount of entanglement as (\ref{inv}) (when discussing the entanglement properties of systems of identical fermions, the relevant group of {\it local transformations} is isomorphic to the group $SU(d)$ of (special) unitary transformations acting on the $d$-dimensional single-particle Hilbert space \cite{ESBL02}).} If the transformation operator commutes with $J_z$, then the transformed state will also be eigenstate of $J_z$ with null eigenvalue, and therefore it will also be an entanglement-invariant state. 

\begin{figure}[h]
	\begin{center}
		\vspace{0.5cm}
		\includegraphics[width=0.35\textwidth]{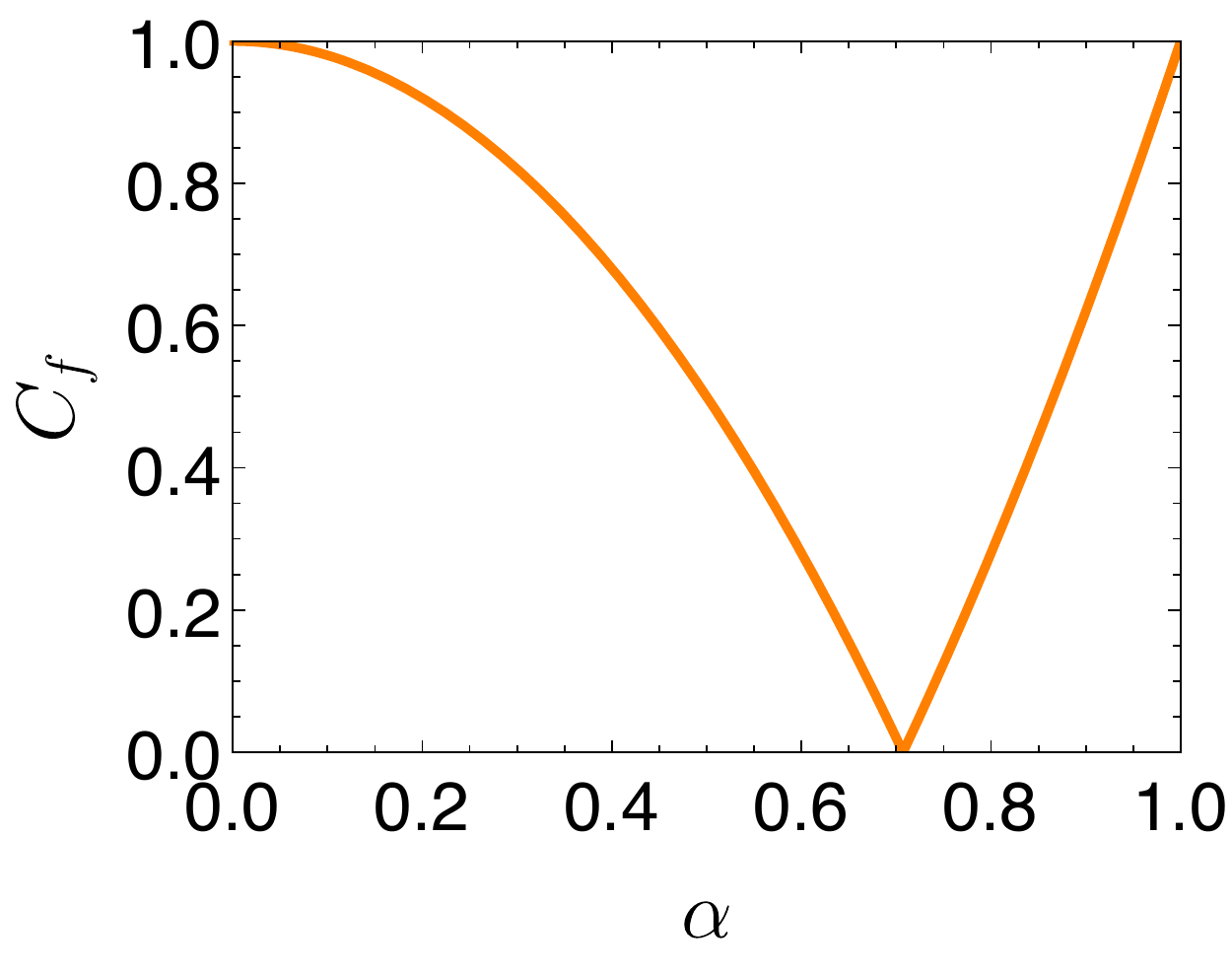}
		\includegraphics[width=0.35\textwidth]{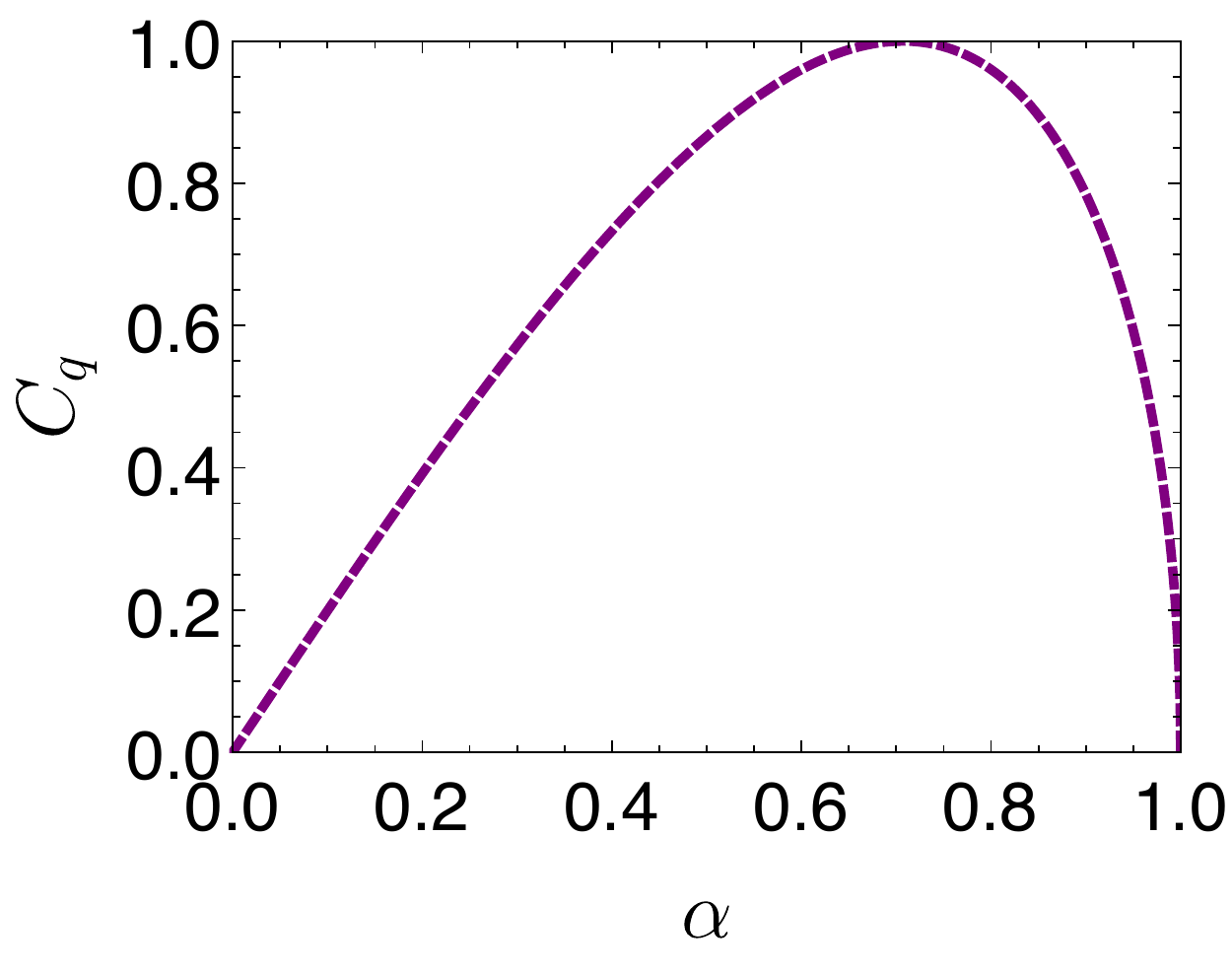}
		\caption[]{\label{fig:Cff_alpha}  Left panel: $C_{f}$ of the state (\ref{inv}) as a function of $\alpha$ (assuming real coefficients). For $\alpha=0$ and $\alpha=1$ the states are maximally entangled (corresponding, respectively, to $\ket{\psi^{-}_{6}}$ and $\ket{\psi^{-}_{3}}$), whereas for $\alpha=1/\sqrt{2}$ the state is a Slater determinant, with zero entanglement. Right panel: $C_{q}$ of the state (\ref{invq}) as a function of $\alpha$ (assuming real coefficients). For $\alpha=0$ and $\alpha=1$ the states are separable (non-entangled), corresponding, respectively, to $\ket{2}$ and $\ket{3}$, whereas for $\alpha=1/\sqrt{2}$ the state is the (maximally entangled) Bell state (\ref{eq:psi_q0}).} 
	\end{center}
\end{figure}

Note that the invariant subspace includes maximally entangled states, particularly, $\ket{\psi^{-}_{3}}$ and $\ket{\psi^{-}_{6}}$. All other amounts of entanglement are attained by varying $\alpha$, as shown in the left panel of Fig. \ref{fig:Cff_alpha}.

For comparison, we will also analyze a system of two-distinguishable qubits in a global environment, considering the initial state $\rho_S(0)=\rho_{qq}(0)=|\psi_{qq}(0)\rangle\langle\psi_{qq}(0)|$, with $|\psi_{qq}(0)\rangle$ a coherent superposition of the elements of the computational basis (\ref{comp}). As before, we take $H_S=\omega_0J_z$ and $\Lambda_S=J_z$, and identify the computational basis with the basis $\{\ket{n}\}$ whose elements satisfy Eq. (\ref{ketn}). Defining
\be
\ket{1}=\ket{00},\ket{2}=\ket{01},
\ket{3}=\ket{10},\ket{4}=\ket{11},
\ee
we thus get
\beq
L_1=-L_4=1;\quad L_2=L_3=0.
\eeq
For the same reasons explained above, also in this (qubit) case the decoherence-free subspace is spanned by the states $\ket{2}$ and $\ket{3}$ (states with null eigenvalues), i.e. (c.f. Eq. (\ref{inv})), 
\beq \label{invq}
|\psi_{qq}(0)\rangle_{\textrm{inv}}=\alpha \ket{2}+\beta \ket{3}.
\eeq
Notice that the (maximally entangled) Bell state
\be\label{eq:psi_q0}
|\phi\rangle=\frac{1}{\sqrt{2}}(|01\rangle + |10\rangle)
\ee
pertains to the invariant subspace, whence two maximally entangled qubits in the state $\ket{\phi}$ can maintain their correlation in spite of the presence of the bath. Other entanglement amounts exhibited by elements of the DFS (Eq. (\ref{invq})) are presented in the right panel of Fig. \ref{fig:Cff_alpha}.

\subsection{Exponential decay}

As follows from the previous paragraphs, if the initial state $|\psi_{S}(0)\rangle$ does not pertain to the corresponding (fermionic or qubit) DFS, the state will in general suffer the effects of the bath exhibiting some degree of decoherence. In order to see the concomitant entanglement and coherence evolution, in this section we thus consider initial states that are orthogonal to the DFS. 

In the fermionic case we will focus on 
\be\label{eq:psi_f0}
|\psi_{ff}(0)\rangle_{24}=\frac{1}{\sqrt{2}}(\ket{\psi^{-}_{2}}+\ket{\psi^{-}_{4}}),
\ee
which is a maximally entangled superposition of states with $m=1$ and $m=-1$, and also on
\be\label{eq:psif15}
|\psi_{ff}(0)\rangle_{15}=\frac{1}{\sqrt{2}}(\ket{\psi^{-}_{1}}+\ket{\psi^{-}_{5}}),
\ee
which is a maximally entangled superposition of states with $m=2$ and $m=-2$. Notice that neither (\ref{eq:psi_f0}) nor (\ref{eq:psif15}) are eigenstates of $\Lambda_S$. From the corresponding $\rho_{ff}(0)$ we determine the evolved matrix and calculate the fermionic concurrence according to Eq. (\ref{concurrencef}), and the coherence as measured by (\ref{coherence}). In addition, since for $T=0$ the joint fermionic system plus environment is in a pure state, the increase in their entanglement can be easily verified
by inspection of the linear entropy $S_L[\rho_{S}]$, quantifying the degree of mixedness in the $S$ subsystem, and given by \cite{Amico}
\begin{equation}\label{SE}
S_L[\rho_{S}]=1-\textrm{Tr}\,\rho_S^2.
\end{equation}

The dynamics of $C_f$, $\mathcal {C}$ and $S_L[\rho_S]$ for the states (\ref{eq:psi_f0}) and (\ref{eq:psif15}) is analyzed in the low temperature regime, resorting to the expressions (\ref{defsc2}). The results are shown and discussed in the Figures below. 

\begin{figure}[!]
	\vspace{0.5cm}
	\begin{center}
		\includegraphics[width=0.35\textwidth]{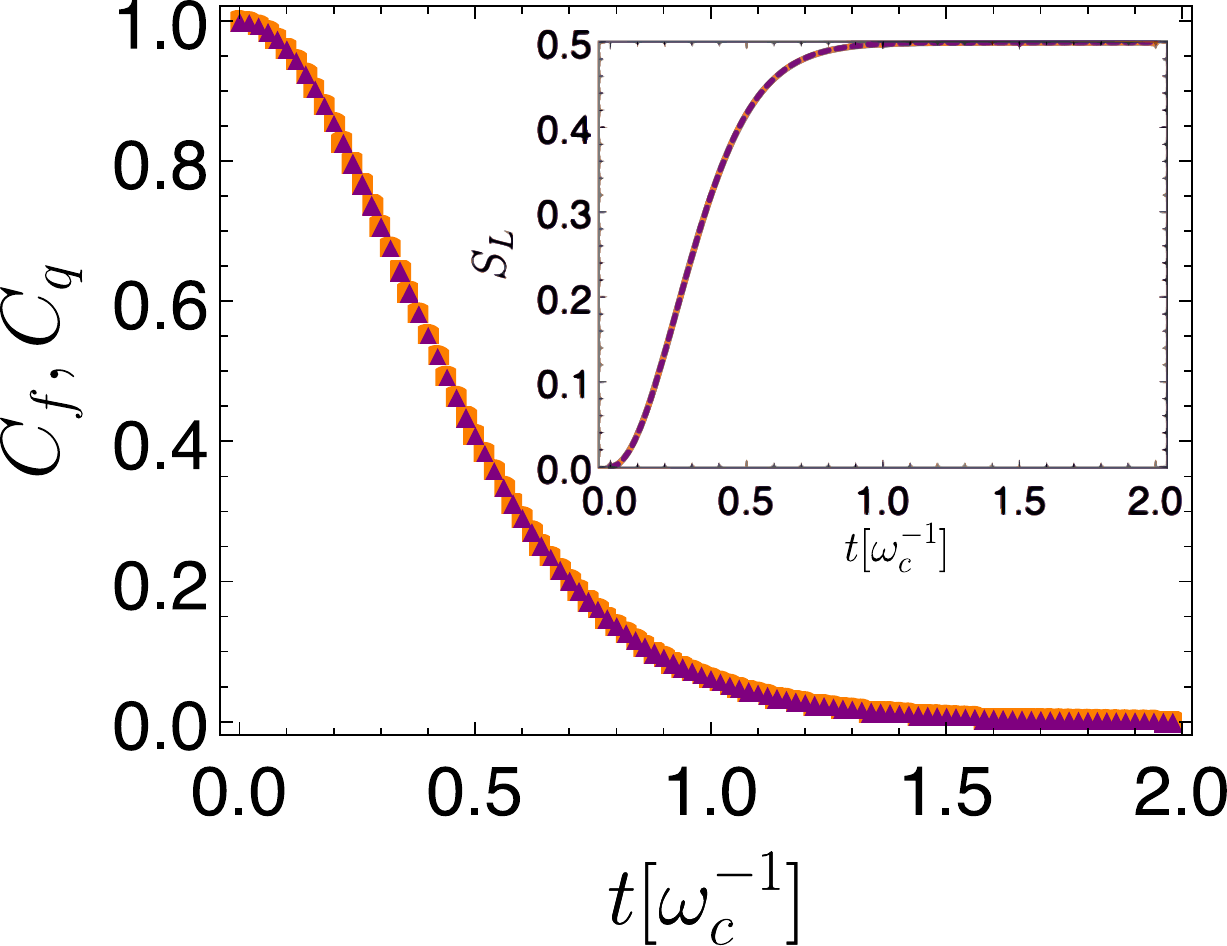}
		\includegraphics[width=0.35\textwidth]{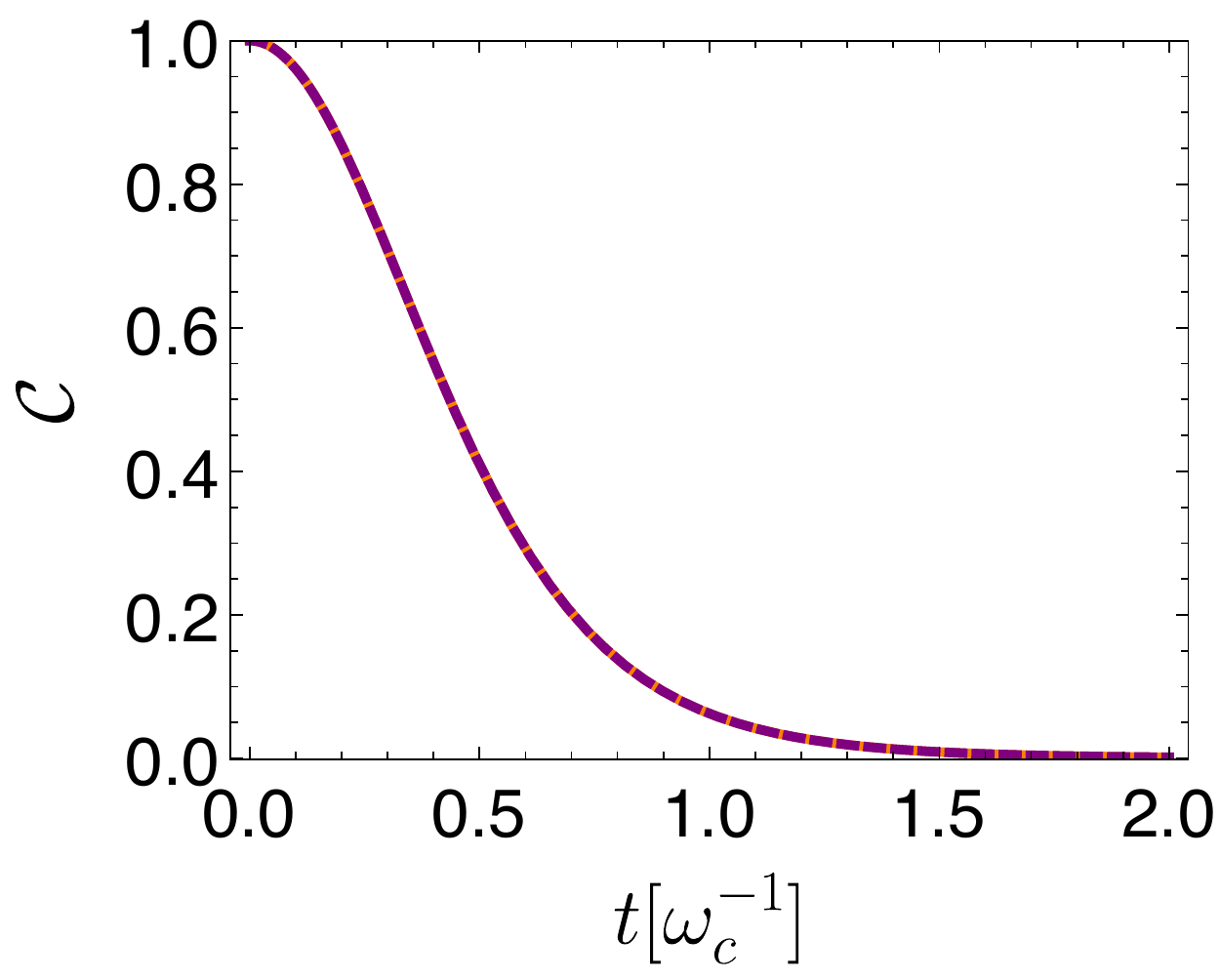}
		\includegraphics[width=0.35\textwidth]{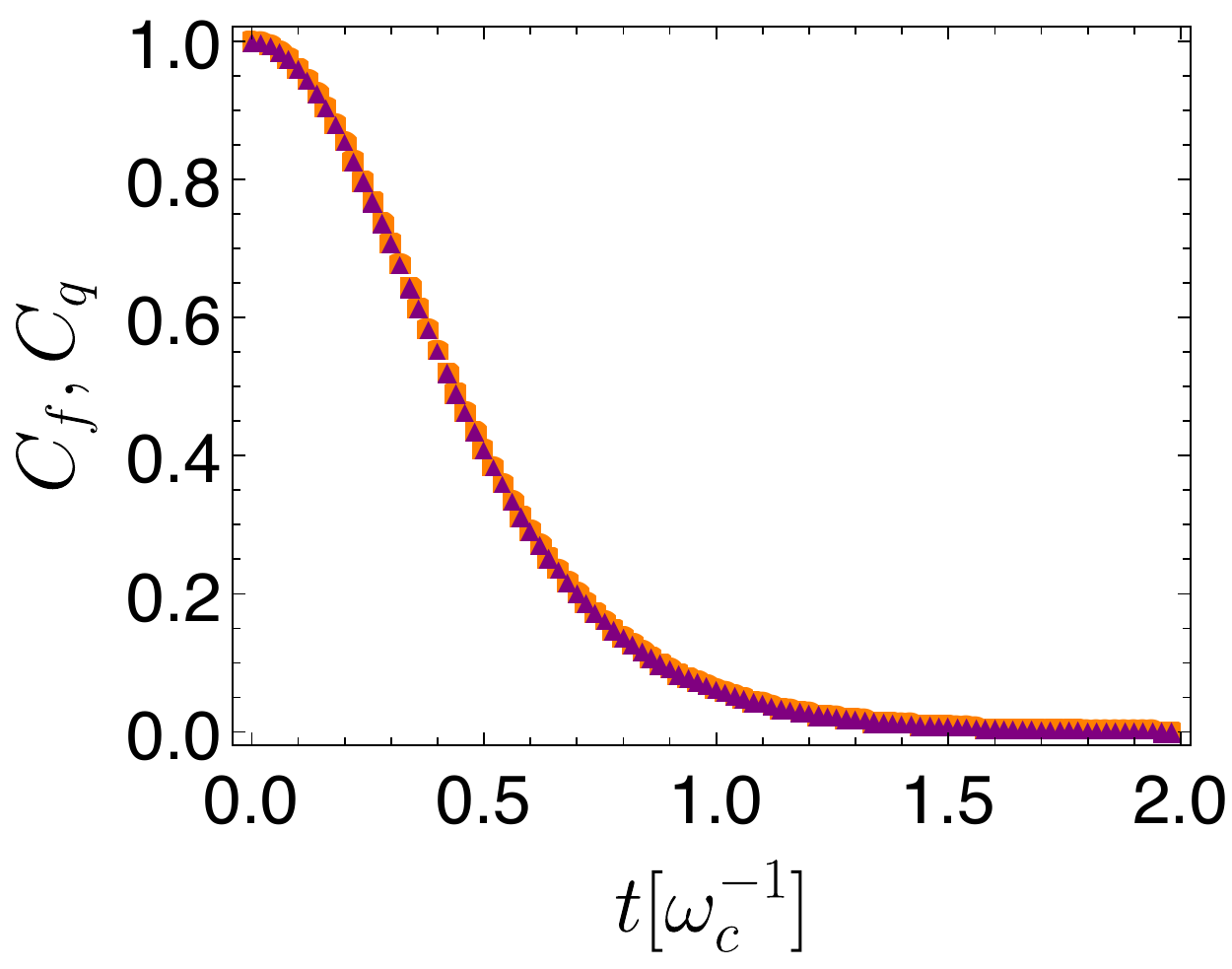}
		\includegraphics[width=0.35\textwidth]{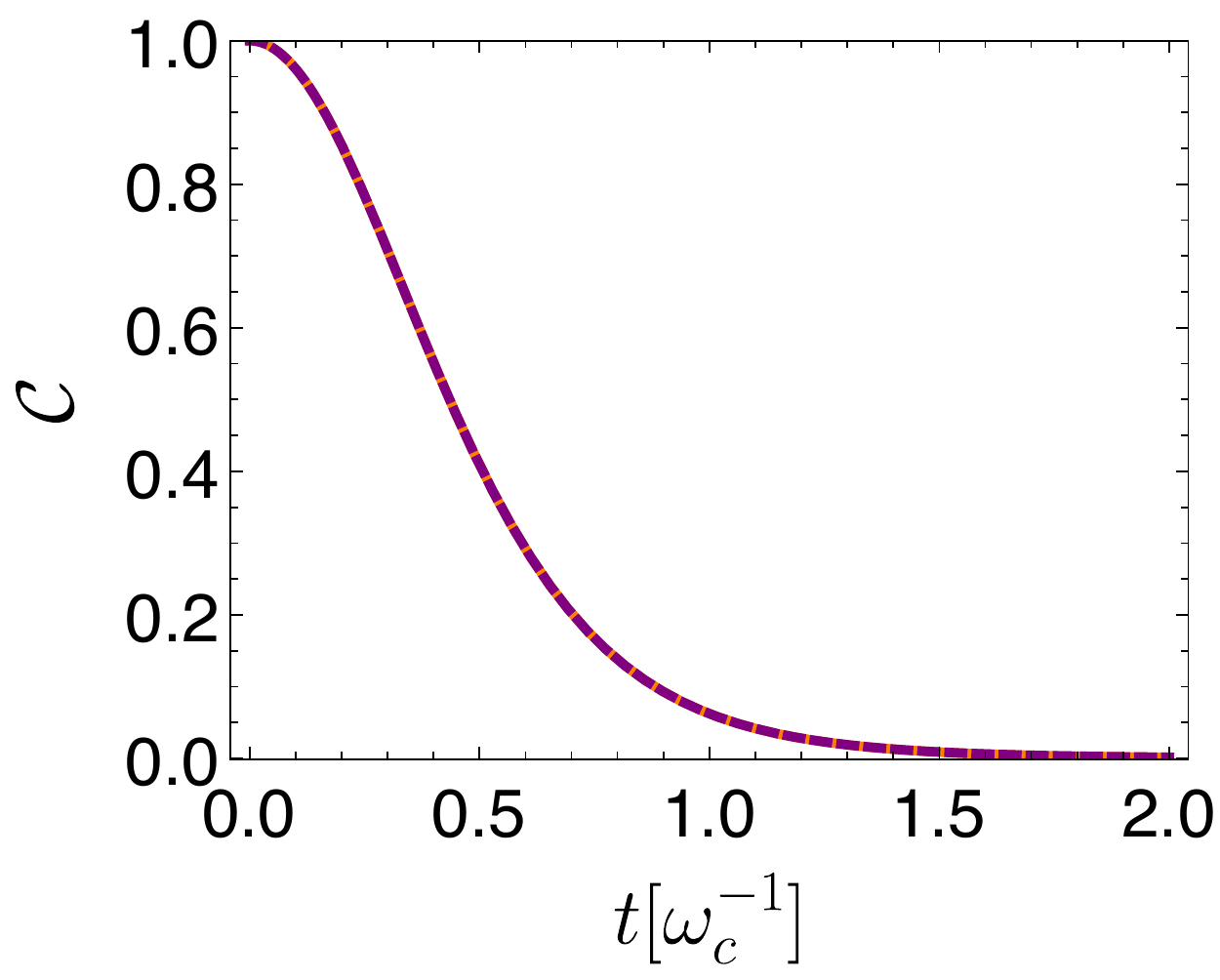}
		\caption[]{\label{fig: state24}(Color online) Left panel: $C_{f}$ (orange curve), and $C_{q}$ (purple curve) for the initial fermion state (\ref{eq:psi_f0}) and the initial qubit state (\ref{eq:psi_qevol}), with $T=0$ (top) and $T/T_c=1/60$ (bottom), as a function of the dimensionless time $t/\omega_c$. Inset: the linear entropy $S_L[\rho_S]$. Right panel: Corresponding coherence measure for the fermionic (orange) and qubit (purple) case with $T=0$ (top) and $T/T_c=1/60$ (bottom).} 
	\end{center}
\end{figure}

As for the qubit system, the state (orthogonal to its corresponding DFS) that will be considered is the (maximally entangled) Bell state
\be\label{eq:psi_qevol}
|\psi_{qq}(0)\rangle_{14}=\frac{1}{\sqrt{2}}(|00\rangle + |11\rangle)=\frac{1}{\sqrt 2}(\ket{1}+\ket{4}).
\ee
\begin{figure}[h]
	\vspace{0.5cm}
	\begin{center}
		\includegraphics[width=0.35\textwidth]{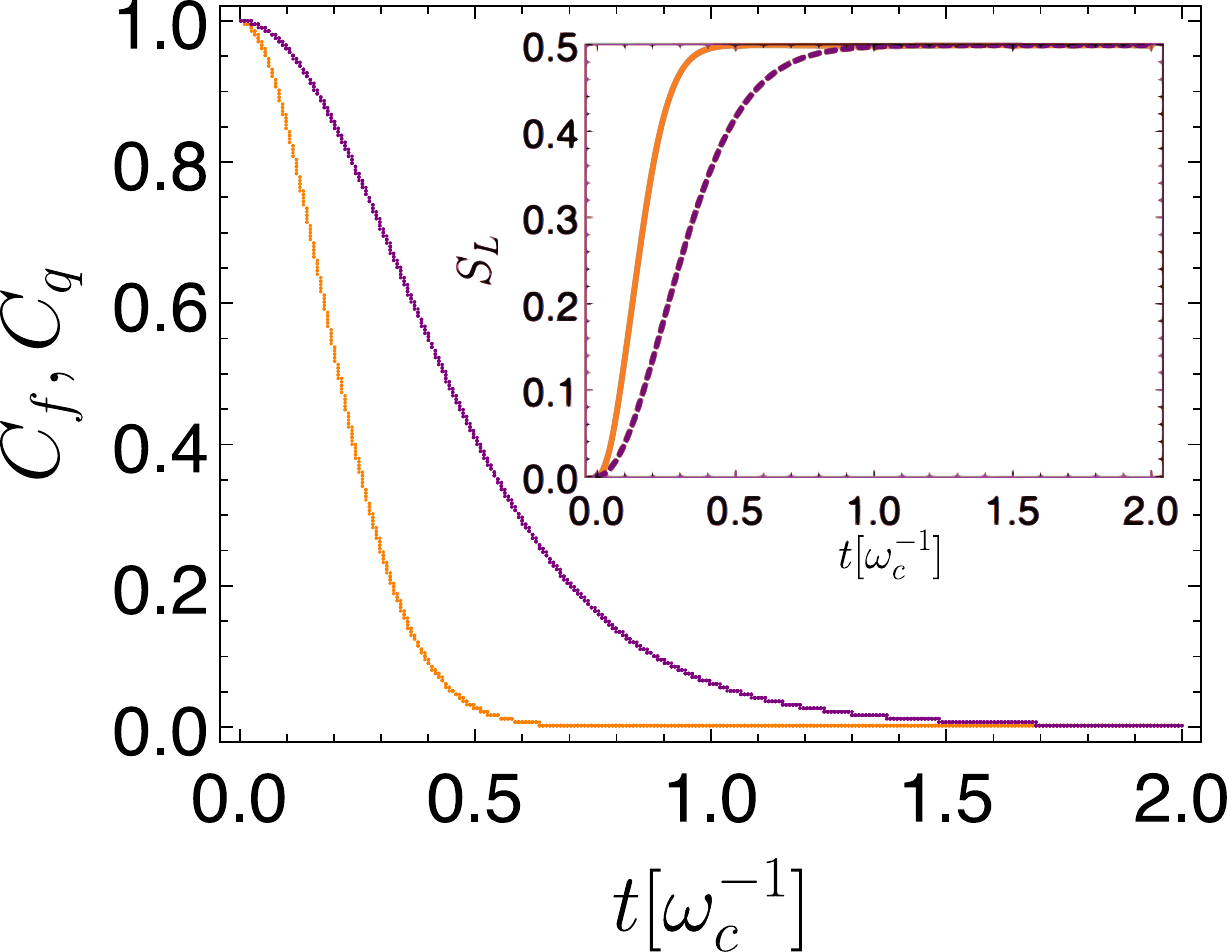}
		\includegraphics[width=0.35\textwidth]{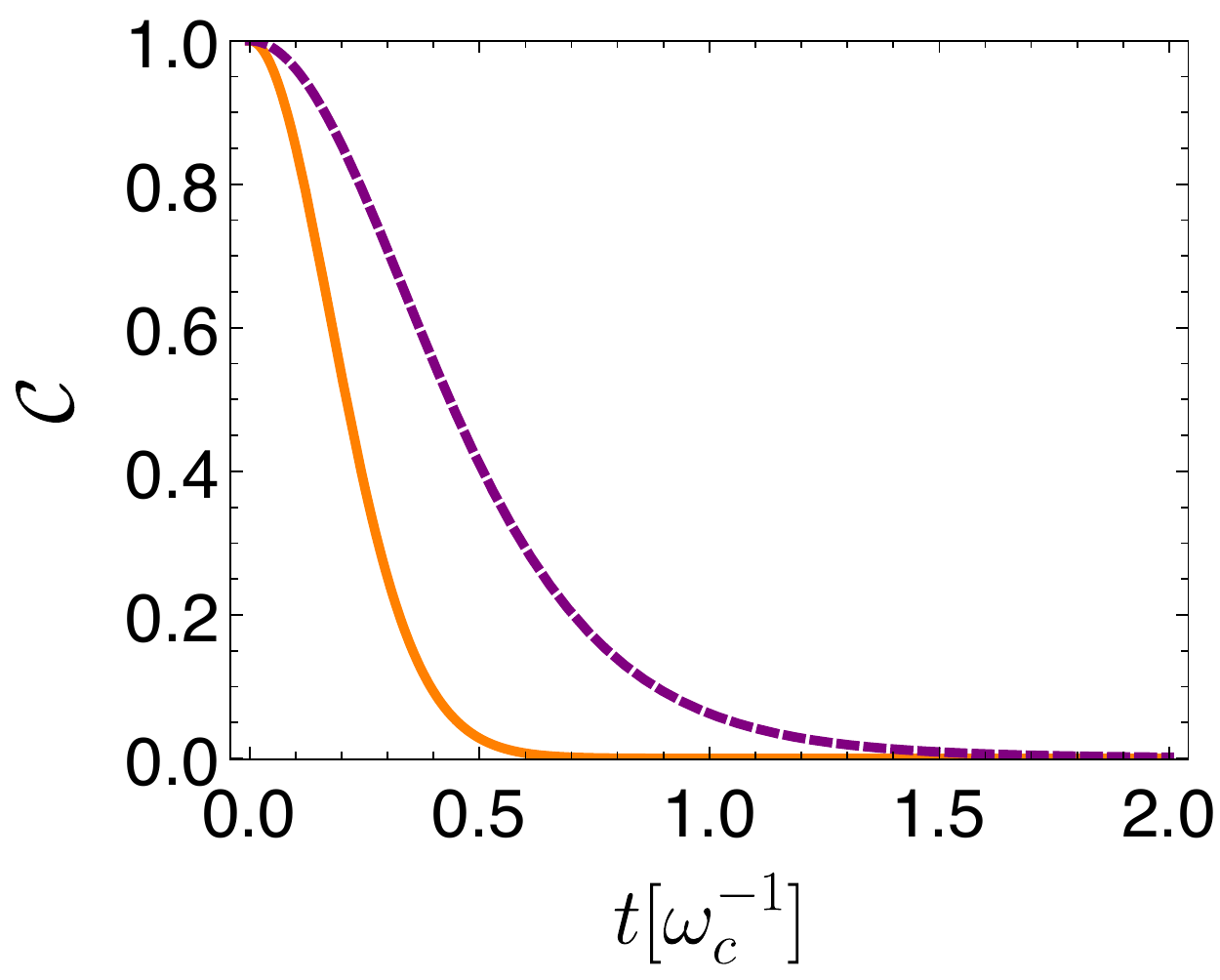}
		\includegraphics[width=0.35\textwidth]{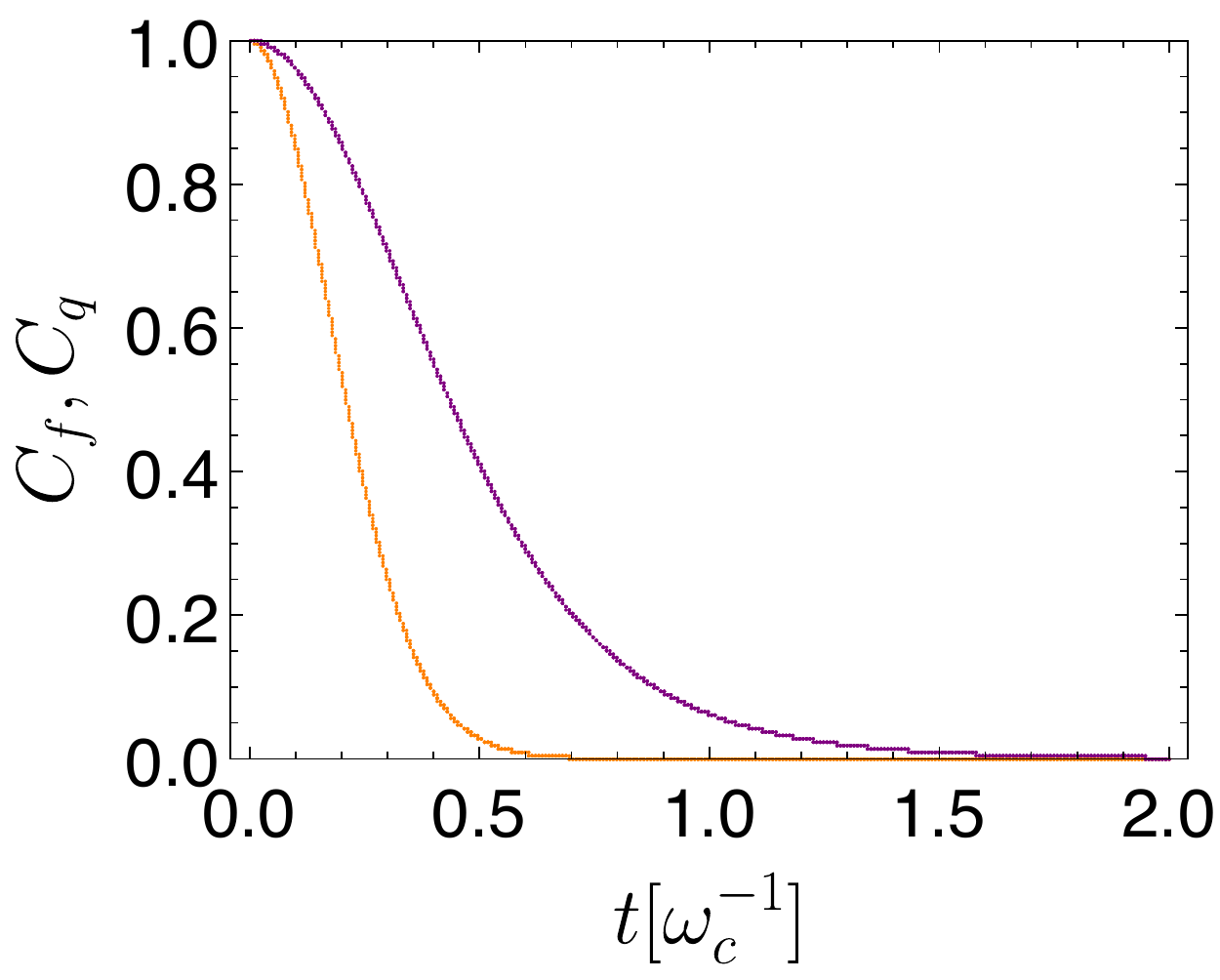}
		\includegraphics[width=0.35\textwidth]{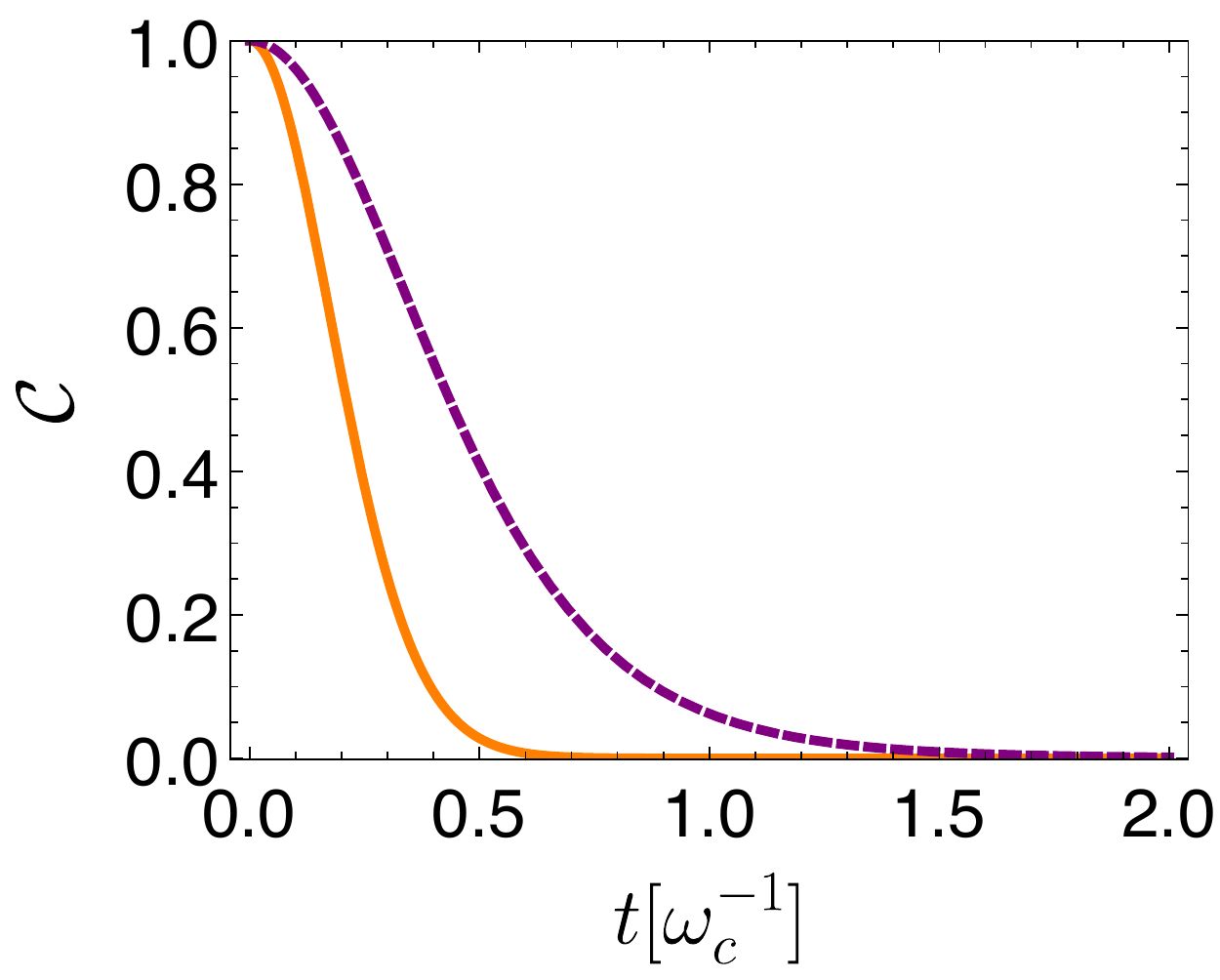}
		\caption[]{\label{fig: state15} (Color online) Left panel: $C_{f}$ (orange curve), and  $C_{q}$ (purple curve) for the initial fermion state (\ref{eq:psif15}) and the initial qubit state (\ref{eq:psi_qevol}), with $T=0$ (top) and $T/T_c=1/60$ (bottom), as a function of the dimensionless time $t/\omega_c$. Inset: the corresponding linear entropy $S_L[\rho_S]$. Right panel: Corresponding coherence measure for the fermionic (orange) and qubit (purple) case with $T=0$ (top) and $T/T_c=1/60$ (bottom).} 
	\end{center}
\end{figure}
The entanglement between the qubits is obtained resorting to the usual concurrence (\ref{concurrence}), whereas the coherence and the linear entropy of the qubits' density matrix are calculated using (\ref{coherence}) and (\ref{SE}), respectively.  

Figure \ref{fig: state24}  shows (left panel) the concurrences $C_{f}$ (orange curve), and  $C_{q}$ (purple curve) for the initial states $|\psi_{ff}(0)\rangle_{24}$ and $|\psi_{qq}(0)\rangle_{14}$, respectively, with $T=0$ (top) and $T/T_c=1/60$ (bottom). The inset (in the zero-temperature case) shows the corresponding linear entropies $S_L[\rho_S]$. In the right panel we show the corresponding evolution of the coherence as measured by $\mathcal C$. In all the four plots the curves superpose; consequently in this case both the fermionic and the qubit system provide the same entanglement and coherence resources throughout the evolution. This results goes in line with that stating that fermionic entanglement between \emph{indistinguishable} fermions, as measured by Eq. (\ref{concurrencef}), is necessary to perform the same tasks that a pair of \emph{distinguishable} entangled qubits with the same amount of entanglement \cite{BouvrieAoP}. Moreover, comparison of the upper and lower panels of the Figure indicates that the behaviour at $T=0$ differs only slightly from that at $T=T_c/60$, which here has been taken as $T=600/60=10$.  

The panels in Figure \ref{fig: state15} show the same quantities as in Fig. \ref{fig: state24} but now referred to the fermion state $|\psi_{ff}(0)\rangle_{15}$ (the qubit state is again $|\psi_{qq}(0)\rangle_{14}$). Clearly the entanglement and the coherence of the fermionic system are more fragile (when compared with the qubit system) under the influence of the environment in this case. Again, there is no appreciable difference in the dynamics for $T=0$ and $T=10$. 

The inset in Figures \ref{fig: state24} and \ref{fig: state15} verifies that as the entanglement between the pair of qubits/fermions decreases, information (as measured by the linear entropy $S_L$) is being loss to the environment (assumed to be in the vacuum, pure, state), or equivalently, the pair as a whole gets entangled with the bath. Notice, however, that the loss of information is not maximal, since $S_L$ saturates before reaching its maximum allowed value, $S_L[\rho_S]_{\max}=1-1/(\textrm{rank}\, \rho_S)$. 

\subsection{Fermionic entanglement sudden death}
\begin{figure}[h]
	\vspace{0.5cm}
	\begin{center}
		\includegraphics[width=0.35\textwidth]{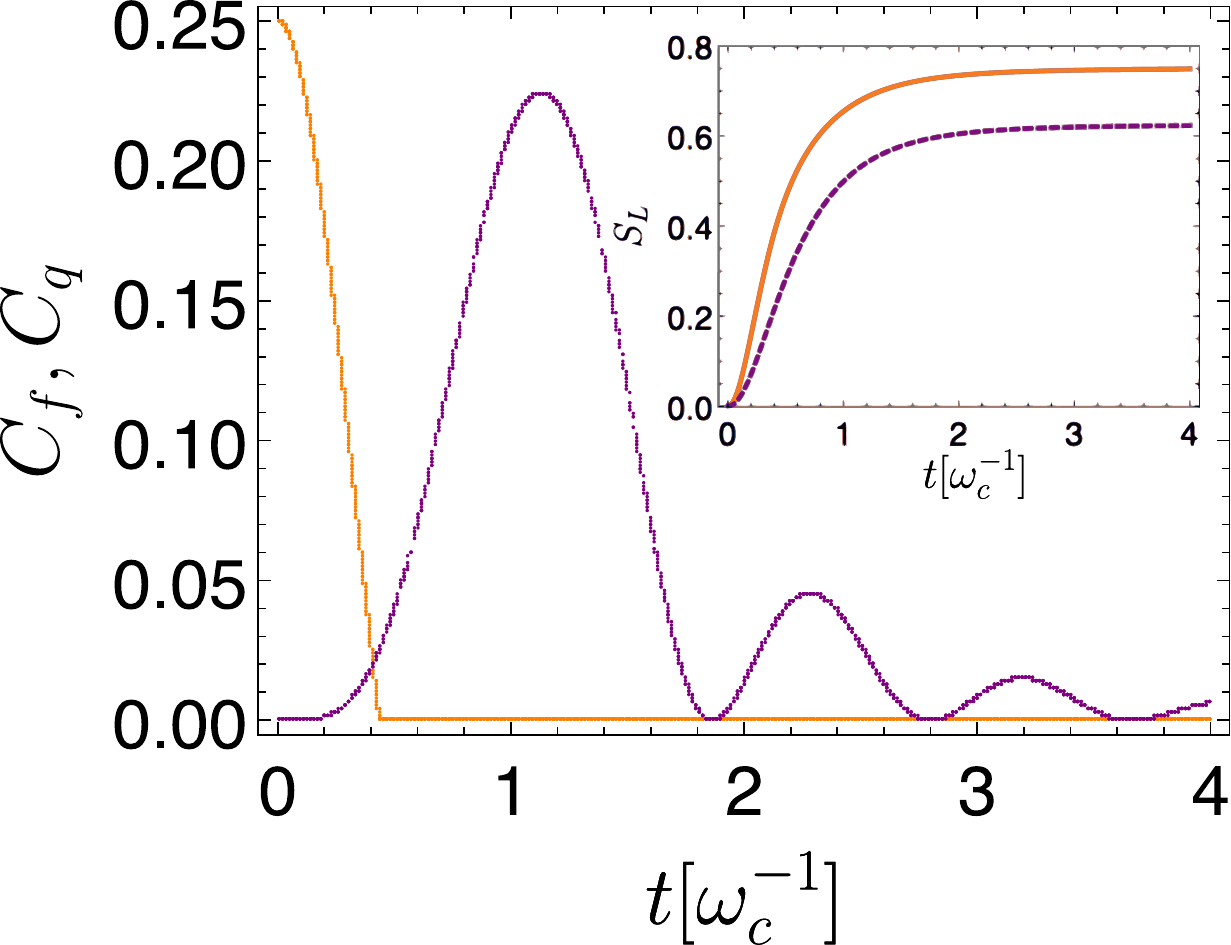}
		\includegraphics[width=0.35\textwidth]{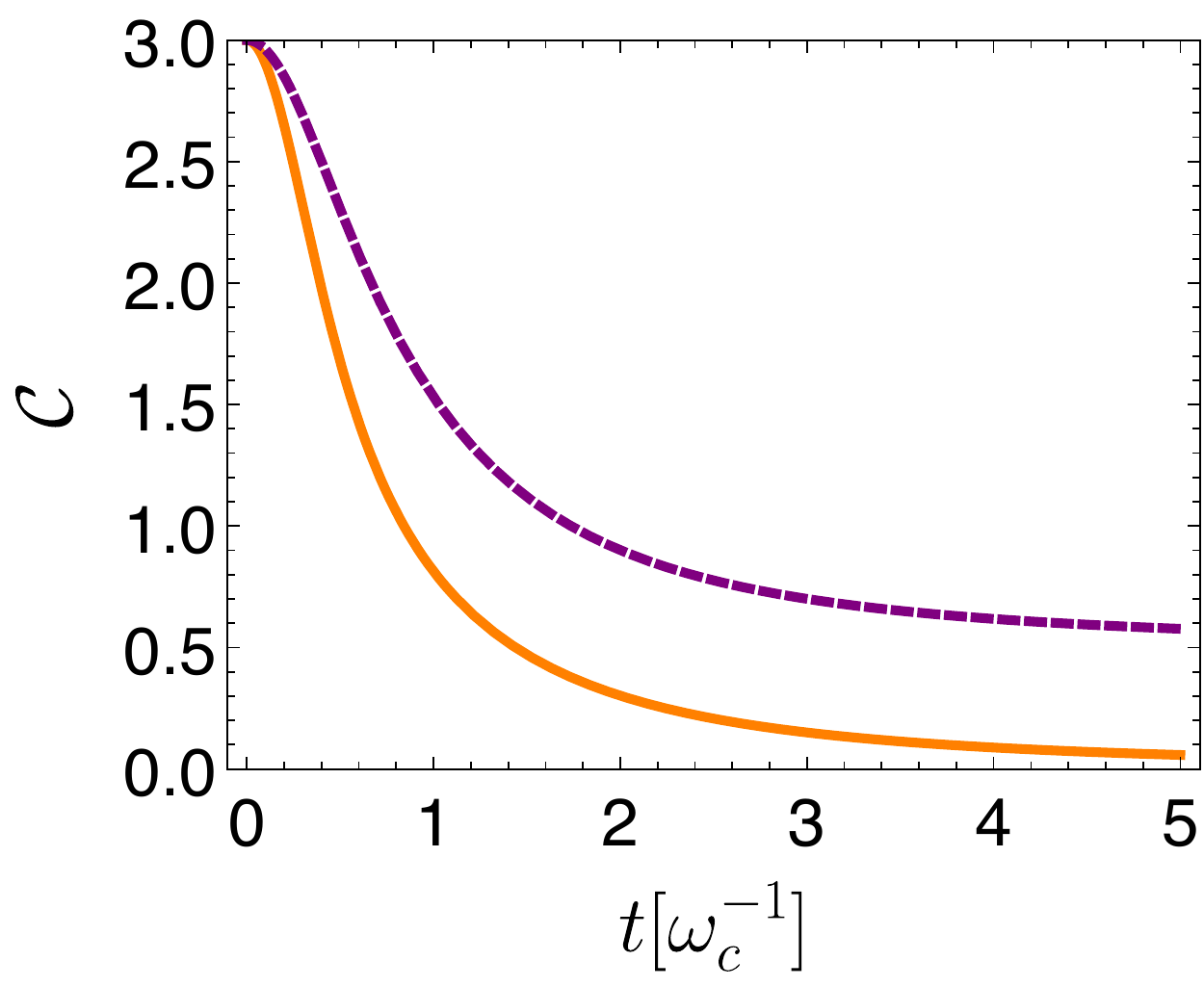}
		\includegraphics[width=0.35\textwidth]{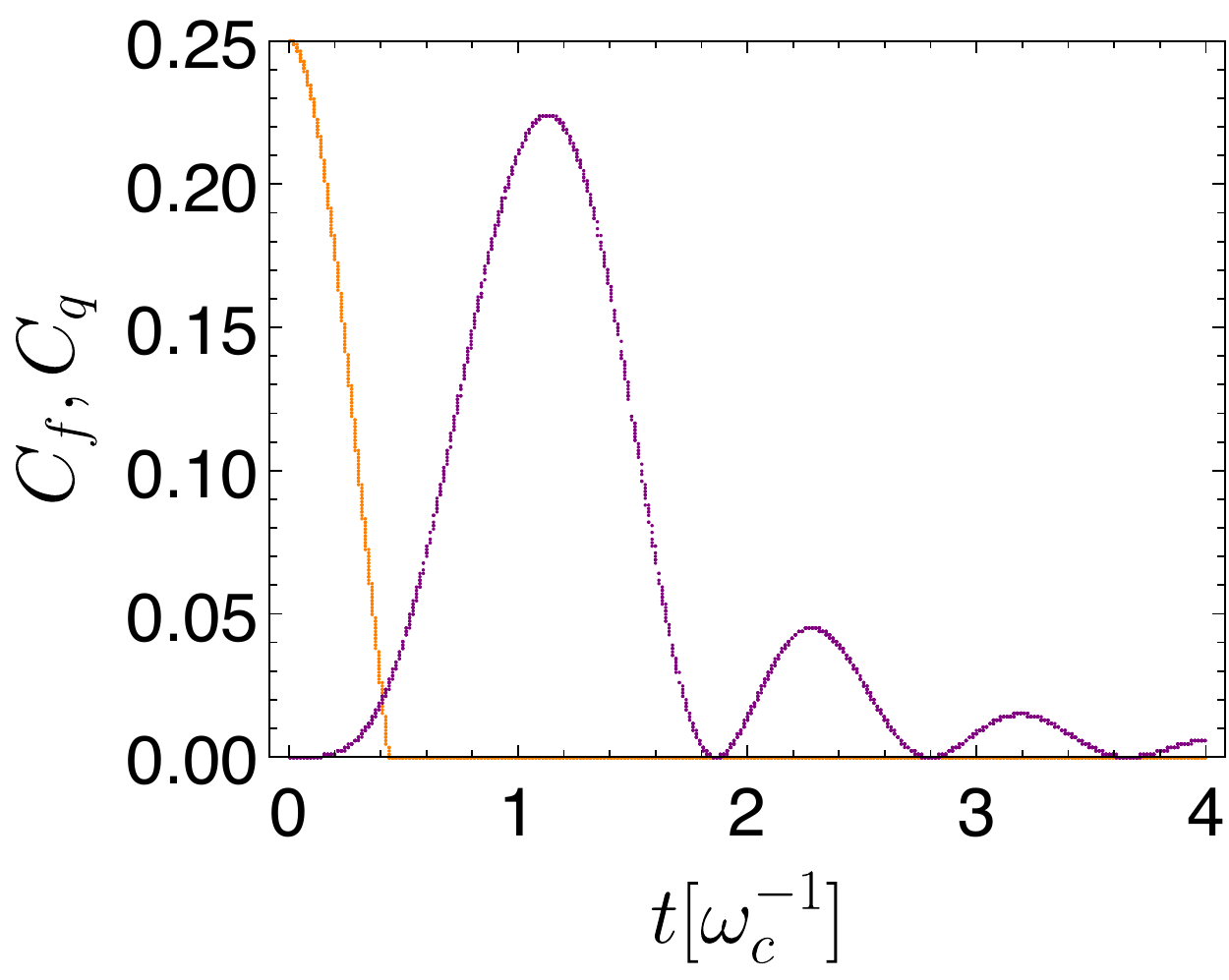}
		\includegraphics[width=0.35\textwidth]{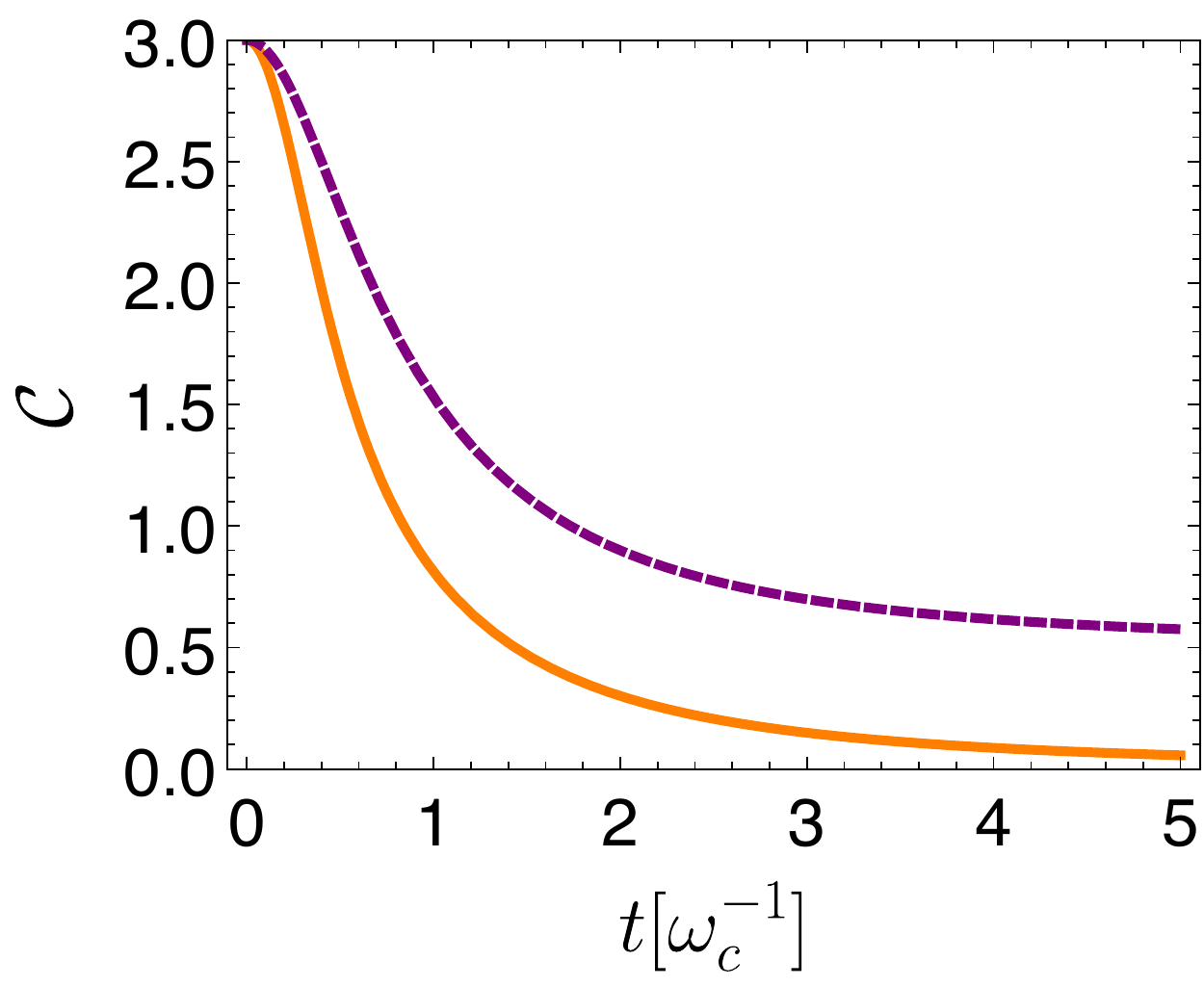}
		\caption[]{\label{fig: crazy}(Color online) Left-panel: $C_{f}$ (orange curve), and  $C_{q}$ (purple curve) for the initial fermion state (\ref{eq:psifc}) and the initial qubit state (\ref{eq:psiqc}), with $T=0$ (top) and $T/T_c=1/60$ (bottom), as a function of the dimensionless time $t/\omega_c$. Inset: the linear entropy $S_L[\rho_S]$. Right-panel: Corresponding coherence measure for the fermionic (orange) and qubit (purple) case.} 
	\end{center}
\end{figure}

Figures \ref{fig: state24} and \ref{fig: state15} involve initial states that are orthogonal to the corresponding (qubit/fermion) DFS, and reflect an asymptotic, monotonous decay both in the entanglement and in the coherence. In order to look for a more varied evolution, we will now consider initial states that have some nonzero overlap with the DFS (notice that when considering an arbitrary initial state this is the more likely situation). We therefore focus now on the initial state 
\be\label{eq:psifc}
|\psi_{ff}(0)\rangle_{1234}=\frac{1}{2}(\ket{\psi^{-}_{1}}+\ket{\psi^{-}_{2}}+\ket{\psi^{-}_{3}}+\ket{\psi^{-}_{4}})
\ee
for the fermions, and on the state
\be\label{eq:psiqc}
|\psi_{qq}(0)\rangle_{1234}=\frac{1}{2}(\ket{1}+\ket{2}+\ket{3}+\ket{4})
\ee
for the qubits. Notably, with these initial conditions the fermionic entanglement rapidly decreases from being maximal, and vanishes abruptly at a finite $t_{\textrm{esd}}$ ---long before the coherence disappears---  therefore exhibiting the phenomenon of \emph{fermionic entanglement sudden death} (see Fig. \ref{fig: crazy}). As follows from the upper-left panel of the figure, at $T=0$ and $t>t_{\textrm{esd}}$ each fermion is disentangled from the other fermion, yet the fermionic pair is entangled with the bath. 

As for the qubit system, notice that the initial state (\ref{eq:psiqc}) is nonentangled. However, as a result of the global interaction, entanglement between the qubits is created until it reaches its maximum value, and from that point on exhibits damped oscillations, vanishing at certain finite times but reviving (or rather, exhibiting \emph{entanglement sudden birth}) until it eventually becomes zero. This type of evolution were pointed out previously for qubits under collective decoherence in \cite{Mazzola2009,Ann2007,Xiang-Ping2010,Ficek2006}. For $T=0$, we observe that the oscillating behaviour of the qubit-qubit entanglement does not affect the monotonous increasing entanglement between the pair of qubits and the environment, indicated by the increase in the linear entropy (see inset), which saturates approximately after the first time $C_q$ vanishes.

As happened in the previous example (Fig. \ref{fig: state15}), the coherence is more robust in the qubit system, yet in this case $\mathcal C$ does not vanishes but tends to a constant value (this is due to the presence of the matrix element $\rho_{23}$, which belongs to the DFS). In its turn, the fermionic coherence decreases more slowly than that shown in Fig. \ref{fig: state15}. Again, no appreciable differences are found in the behaviour for $T=0$ and $T=10$.   

\begin{figure}[h]
	\vspace{0.5cm}
	\begin{center}
		\includegraphics[width=0.35\textwidth]{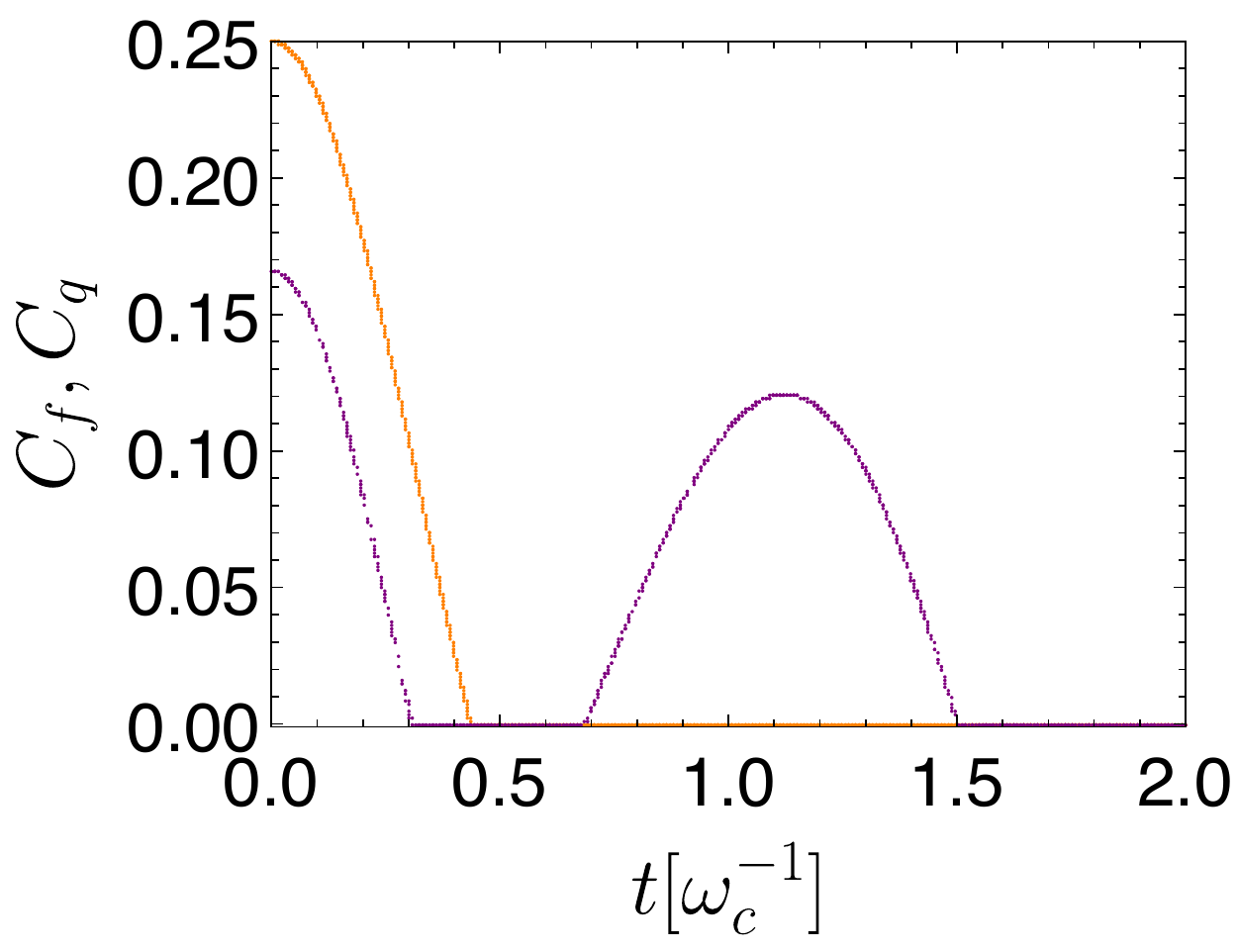}
		\includegraphics[width=0.35\textwidth]{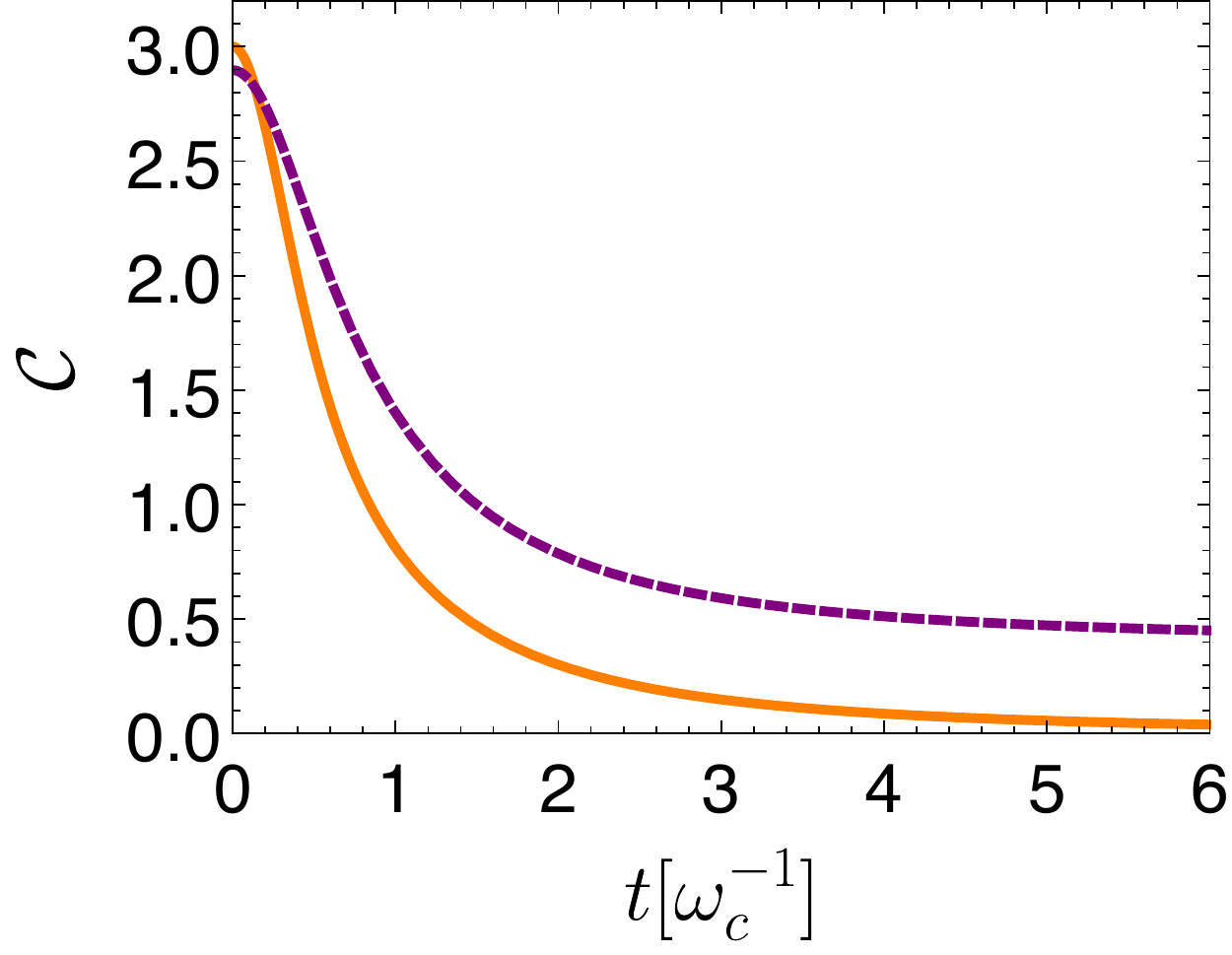}
		\caption[]{\label{fig: crazybis}(Color online) Left: $C_{f}$ (orange curve), and  $C_{q}$ (purple curve) for the initial fermion state (\ref{eq:psifc}) and the initial qubit state (\ref{eq:psiq2}), with $T=10$, as a function of the dimensionless time $t/\omega_c$. Right: Corresponding coherence measure for the fermionic (orange) and qubit (purple) case.} 
	\end{center}
\end{figure}
As a final example, we will consider the fermionic initial state (\ref{eq:psifc}) and the initial qubit (entangled) state
\be\label{eq:psiq2}
|\psi_{qq}(0)\rangle_{123-4}=\sqrt{0.2}(\ket{1}+\ket{2}+\ket{3})+\sqrt{0.4}\ket{4},
\ee
and compare the dynamics of the corresponding concurrences and coherences for $T=10$. The resulting curves (indistinguishable from those at $T=0$) are depicted in Fig. \ref{fig: crazybis}. Interestingly, in this case the qubit-qubit system exhibits entanglement sudden death and entanglement sudden birth, while its coherence decreases and tends to a constant value. 

\section{Concluding remarks\label{sec:conclusions}}

A better understanding of the entanglement and coherence evolution under different quantum channels is a suitable way to reach a more complete view of the potentialities of a quantum system in quantum information processing. In the present work, we have investigated the non-dissipative Markovian evolution of entanglement and coherence in the simplest fermionic (pure) states that exhibit the phenomenon of fermionic entanglement, and compared its evolution with that of a two-distinguishable-qubit system collectively coupled to the environment. As for the latter, we have considered a thermal bosonic reservoir coupled to the central (fermionic/qubit) system. 

In spite of the simplicity of the fermionic system (equivalent to two $3/2$-spin fermions), we have found interesting results regarding the existence of decoherence-free subspaces, and the emergence of entanglement sudden death.  

If the initial fermionic state belongs to the DFS, the state will remain unaffected by the interaction with the bath throughout the evolution. The identification of these kind of subspaces, generated by collective coupling, constitutes a possible solution to the decoherence problem in QIP \cite{Divincenzo2016,Duan1998,Duan1998v2}. Thus, the previous results indicate a possible way to avoid decoherence in the indistinguishable-party case via collective coupling. 

If the initial fermionic state does not belong to the invariant subspace, the entanglement and coherence evolve in time. To contrast their evolution with that corresponding to the pair of distinguishable parties, we have considered a 2-qubit system subject to the same environment. Figs. \ref{fig: state24} and \ref{fig: state15} indicate that both the entanglement and the coherence in the fermionic system are no more robust under collective decoherence that those in the qubit system, when the initial states are orthogonal to the corresponding (fermion or qubit) DFS. 

On the other hand, by considering initial states that have non-zero overlap with the elements of the corresponding DFS, we find a much richer evolution for both the qubit-qubit and the fermion-fermion entanglement. In particular, we showed that the fermionic entanglement exhibits entanglement sudden death, and no revival is observed, whereas the qubit entanglement can exhibit a damped oscillating behaviour, and also entanglement sudden death and entanglement sudden birth.   

\begin{acknowledgements}
\noindent D.B. and A.P.M. acknowledge support via Grant No. GRFT-2018 MINCYT-Córdoba, as well as the Argentinian agencies SeCyT-UNC and CONICET for their financial support.. D. B. has a fellowship from CONICET. A.V.H. acknowledges financial support from DGAPA, UNAM through project PAPIIT IN113720.
\end{acknowledgements}

{}
\end{document}